\documentclass[aps,prl,groupedaddress,twocolumn,10pt]{revtex4-1}
\usepackage{graphicx}
\usepackage{epstopdf}
\usepackage[english]{babel}
\usepackage[T1]{fontenc}
\usepackage{verbatim}
\usepackage{float}

\usepackage{makecell}
\usepackage{color}
\usepackage{multirow}
\usepackage{booktabs}
\usepackage{amsmath}
\usepackage{amssymb}
\usepackage{amsthm}
\usepackage{bbm}
\usepackage{hyperref}
\usepackage[T1]{fontenc}
\usepackage[usenames,dvipsnames]{xcolor}
\hypersetup{colorlinks=true, linkcolor=blue, urlcolor=blue!50!black, citecolor=blue}

\usepackage[normalem]{ulem}

\renewcommand{\imath}[0]{\mathsf{i}}

\usepackage[nodayofweek,level]{datetime}

\begin{document}

\title{Wrapping dynamics and full uptake conditions for  nonspherical active nanoparticles}
\author{Ke Xiao}
\email{xiaoke@ucas.ac.cn}
\affiliation{Wenzhou Institute, University of Chinese Academy of Sciences, Wenzhou 325016, People's Republic of China}
\affiliation{Department of Physics, College of Physical Science and Technology, Xiamen University, Xiamen 361005, People's Republic of China}
\author{Rui Ma}
\affiliation{Fujian Provincial Key Lab for Soft Functional Materials Research, Research Institute for Biomimetics and Soft Matter, Department of Physics, College of Physical Science and Technology, Xiamen University, Xiamen 361005, People's Republic of China}
\author{Chen-Xu Wu}
\email{cxwu@xmu.edu.cn}
\affiliation{Fujian Provincial Key Lab for Soft Functional Materials Research, Research Institute for Biomimetics and Soft Matter, Department of Physics, College of Physical Science and Technology, Xiamen University, Xiamen 361005, People's Republic of China}

\begin{abstract}
The cellular uptake of self-propelled nanoparticles (NPs) or viruses, usually nonspherical, by cell membrane is crucial in may biological processes. In this study, using Onsager variational principle, we obtain a general wrapping equation for nonspherical self-propelled nanoparticles. Two analytical critical conditions are theoretically derived, one for the continuous full uptake of prolate particles and the other for snapthrough full wrapping of oblate particles. They capture considerably well the full uptake critical boundaries in the phase diagrams constructed in terms of active force, aspect ratio, adhesion energy density, and membrane tension based on numerical calculations. It is found that enhancing activity (active force), reducing effective dynamic viscosity, increasing adhesion energy density, and decreasing membrane tension, can significantly improve the wrapping efficiency for the self-propelled particles. These results elucidate some of the previous specific investigations conclusively and may offer novel possibilities for designing an effective active NP-based vehicle for controlled drug delivery.
\end{abstract}
\date{\today}

\maketitle

The lipid bilayer plasma membrane, a physical barrier defining organelles of cells and plenty of their surrounding environment, plays a crucial role for a spectrum of biological processes~\cite{B.Alberts2005}. Examples range from the transduction of biochemical signal and the intake of nutrients~\cite{F.Frey2019PRE} to budding and fission~\cite{G.V.Meer2004,J.T.Groves2010}, and endocytosis of viruses, pathogens, and particles~\cite{G.Bao2005,S.Zhang2015}.
The engulfing of a particle or virus (pathogen) by a plasma membrane is a widely encountered phenomenon in endocytosis processes, including inter- and intracellular transport~\cite{S.Behzadi2017}, delivering therapeutic agents enveloped by nanoparticles into tumor cells~\cite{D.Peer2007,W.Rao2015,Y.Min2015}, and virus infection~\cite{N.Kol2007,R.F.Bruinsma2021,C.B.Jackson2022}.
Especially, cellular uptake, which involves the interaction between cell membrane and nanoparticles (NPs) or viruses, is an essential step for a wide range of healthy and disease-related processes~\cite{R.F.Bruinsma2021}.

Over the past two decades, considerable efforts using experiment, theoretical modeling, and numerical simulation have been devoted to characterizing how the physical parameters, including the particle size ~\cite{Deserno2002,Deserno2003,M.Deserno2004,S.Zhang2009,B.D.Chithrani2006,J.Agudo2015,C.Contini2020}, shape~\cite{F.Frey2019,K.Yang2010,Z.Shen2019}, elastic properties of the particle~\cite{X.Yi2011,J.C.Shillcock2005,X.Yi2014,A.Verma2010,X.Ma2021}, ligand/receptor density~\cite{H.Yuan2010PRL,H.Yuan2010,T.Wiegand2020}, as well as the mechanical properties of the membrane~\cite{J.Agudo-Canalejo2015, H.T.Spanke2020}, affect the invading behaviors.
Though the cellular uptake of passive particles via endocytic process has been studied extensively, little work has been done on the active entry of self-propelled bacterial pathogens.
To name a few examples, it has been found that some cytosolic bacteria such as \textit{Rickettsia rickettsii} are able to produce active force to facilitate their mobility by forming actin tails~\cite{P.M.Colonne2016}, and \textit{Listeria monocytogenes} can generate active force to push out a tube-like protuberance from the plasma membrane by hijacking the actin polymerization-depolymerization apparatus of their host~\cite{J.A.Theriot1992,J.R.Robbins1999,T.Chakraborty1999,F.E.Ortega2019,G.C.Dowd2020}.
How the active force of these self-propelled agents affects the engulfing dynamics at the cell membrane remains to be elucidated.

Recently, many model systems by using lipid vesicles to encapsulate natural swimmers (\textit{Escherichia coli} bacteria, \textit{Bacillus subtilis} bacteria, etc.) or artificial microswimmers (synthetic Janus particles) have been developed to study the active membrane behaviors in vitro~\cite{H.R.Vutukuri2020,S.C.Takatori2020,C.Wang2019,Y.Li2019,M.S.E.Peterson2021,L.LeNagarda2022}.
Such systems are out-of-equilibrium and hence give rise to many intriguing behaviors, such as membrane fluctuations and large deformations~\cite{H.R.Vutukuri2020,S.C.Takatori2020}, shape transformations~\cite{C.Wang2019,Y.Li2019,M.S.E.Peterson2021}, and even deformation of lipid vesicles into flagellated swimmers~\cite{L.LeNagarda2022}.
Therefore, in biology, the specific interactions between vesicles and bacteria or artificial self-propelled particles plays a key role in designing active matter systems~\cite{A.T.Brownet2016}. Besides, the wrapping dynamics of particles by cell membranes, which is important for understanding the cellular uptake, has been carried out experimentally, theoretically, and numerically. In the limit of low membrane tension and weak reversible adhesion, Spanke \textit{et. al.}~\cite{H.T.Spanke2022} experimentally investigated how the spontaneous wrapping dynamics of micron-sized particles by giant unilamellar vesicles changes with the adhesion energy.
By combining computer simulations and theoretical analysis, the cellular uptake of active particles in the absence of membrane tension was studied~\cite{P.Chen2020}, and the deterministic and stochastic uptake dynamics of passive nanoparticles with different geometries were also reported~\cite{F.Frey2019}. Understanding such effect of forces and membrane properties (adhesion energy density and membrane tension) on the dynamics of cellular uptake are critical to designing efficient strategies for potential biomedical applications, including drug/gene delivery~\cite{Panyam2003,Xu2018,Wang2019}, cell operation and manipulation~\cite{M.Medina2018,Xie2019}, and bioimaging/sensing~\cite{R.Weissleder2006,D.Peer2007}.
In addition, in reality many pathogens and viruses are nonspherical~\cite{C.Hulo2011}, such as egg-shaped \textit{malaria parasite}~\cite{S.Dasgupta2014} and cylindrical \textit{Listeria monocytogenes}, indicating the significance of probing the wrapping dynamics of nonspherical particles for the entry of certain pathogens into cells.

To better understand the cellular uptake dynamics of a nonspherical self-propelled nanoparticle by a plasma membrane, a detailed and comprehensive investigation of how the wrapping time depends on the active force, the particle's aspect ratio, the viscosity, and the membrane properties (adhesion energy density and membrane tension) is needed.
Here in this paper, the wrapping dynamics is studied by employing the Onsager variational principle~\cite{OnsagerI1931,OnsagerII1931} for out-of-equilibrium systems. Our results show that active force, low effective dynamic viscosity, strong adhesion force, and small membrane tension play a positive role in the wrapping time, indicating that it can be manipulated by changing the activity and the aspect ratio of the particles, the viscosity, and the properties of the membrane such as adhesion energy density and membrane tension. The physical insights gained from this work could clarify the mechanism of wrapping dynamics of nonspherical active nanoparticles.

\begin{figure}[htp]
  \includegraphics[width=\linewidth,keepaspectratio]{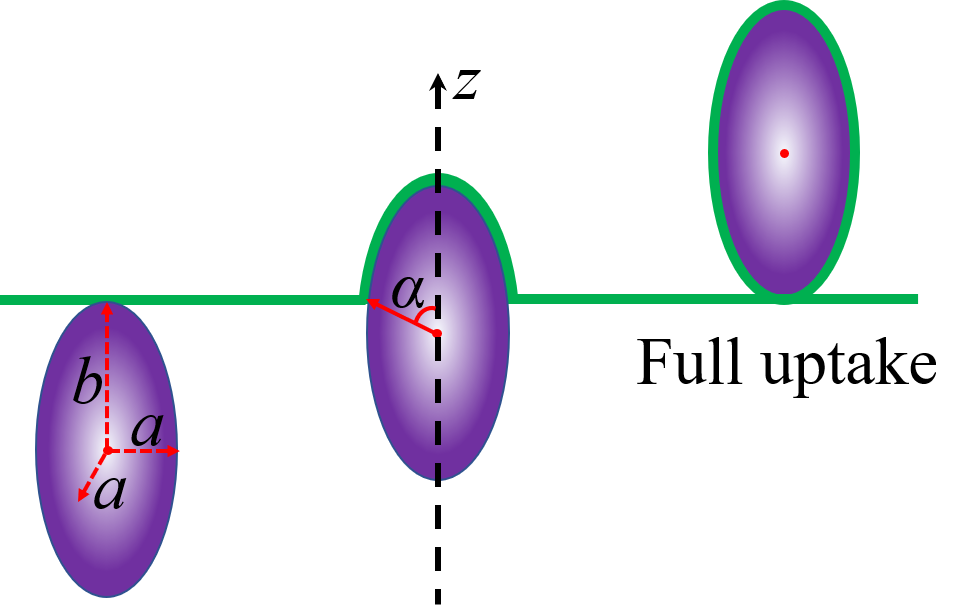}
  \caption{(Color online) Schematic depicting wrapping phases of an active particle from nonwrapping to partial wrapping and to full uptake. \label{uptakeprocess}}
\end{figure}

We model a self-propelled invader as an axis-symmetric ellipsoid (prolate or oblate spheroid) with its principle rotational axis orthogonal to the flat membrane, as shown in Fig.~\ref{uptakeprocess}, where
$a$ and $b$ denote the semi-axes perpendicular to and along the principle rotational axis, respectively. The geometry of the particle is parameterized by the aspect ratio $e=b/a$, with $e>1$ for a prolate ellipsoid, and $e<1$ for an oblate one. In practice, it is convenient to write the area element in terms of polar angle: $
dA=2\pi a^2 \sin\theta\sqrt{\cos^2\theta+e^2\sin^2\theta}~d\theta.$
To model the uptake dynamics of an active particle, we have to write down the total free energy of the system. Following the classical Canham-Helfrich continuum model~\cite{Helfrich1973,F.Julicher1994,M.Deserno2004}, such an energy is given by
\begin{align}
E_{\mathrm{tot}}=\int_{A_{\rm mem}}\frac{\kappa}{2}(2H)^2 dA +\sigma\Delta A - \int_{A_{\rm ad}}\omega~dA - fZ,\label{HelfrichFE}
\end{align}
where the elastic energy of the membrane, the adhesion energy between the particle and the membrane, and the work done by the active particle are taken into account.
Here the first term, with $\kappa$ the bending rigidity and $H$ the local mean curvature, denotes the bending energy of the membrane
\begin{align}
E_{\rm bend}=&\int_0^{\alpha}\pi\kappa e^2\sin\theta\frac{[2+(e^2-1)\sin^2\theta]^2}{[1+(e^2-1)\sin^2\theta]^3}\times \notag\\ &\sqrt{\cos^2\theta+e^2\sin^2\theta}~d\theta,\label{E_bend}
\end{align}
an integral over the contact area between the membrane and the particle. The second term of Eq.~(\ref{HelfrichFE}) is contributed by the surface tension:
\begin{align}
E_{\rm ten}=&\int_0^{\alpha}2\pi\sigma a^2 \sin\theta \biggl[1-\frac{\cos\theta}{\sqrt{\cos^2\theta+e^2\sin^2\theta}}\biggr]\times \notag\\ &\sqrt{\cos^2\theta+e^2\sin^2\theta}~d\theta.\label{E_tension}
\end{align}
The third term of Eq.~(\ref{HelfrichFE}) representing the gain in adhesive energy, characterized by a negative adhesive energy $-\omega$ per unit area, can be written as an integral over the wrapping area. The last term of Eq.~(\ref{HelfrichFE}) arises from work done by the active force $f$ acting on the particle, which is calculated as $E_{\rm f}=-fae(1-{\rm cos}\alpha)$.

As the particle is being engulfed by the membrane, it has been found that it is the friction force near the membrane-particle contact line with its circumference given by $L(\alpha)=2\pi a\sin\alpha$ that largely dissipates the energy~\cite{H.T.Spanke2022}. In the limit of low Reynolds number, the
\begin{figure*}[htp]
  \includegraphics[width=\linewidth,keepaspectratio]{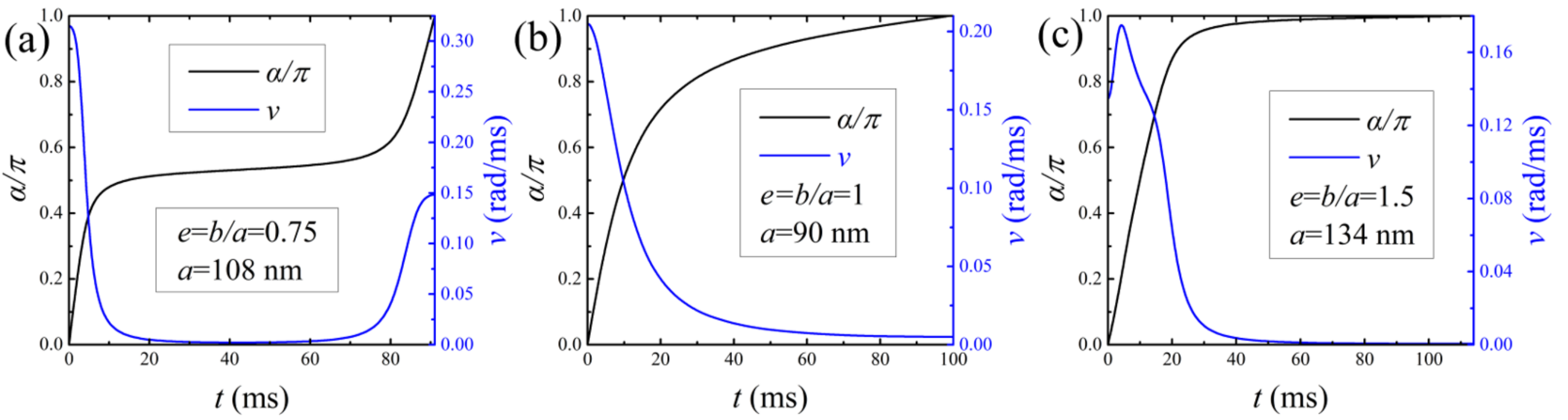}
  \caption{(Color online)  The engulfing angle $\alpha/\pi$ and the wrapping velocity $v$ of an active particle with different aspect ratios (a) $e=0.75$, (b) $e=1$, and (c) $e=1.5$, where the bending rigidity, the adhesion energy density, the tension modulus, and the effective dynamic viscosity are set as $25~k_BT$, $0.044~{\rm mJ/m^2}$, $0.9\times10^{-5}~{\rm N/m}$, and $1~{\rm Pa\cdot s}$, respectively.} \label{tVsR}
\end{figure*}
 dissipation function reads
\begin{align}
\Phi=\pi\eta a^3\sin\alpha (\cos^2\theta+e^2\sin^2\theta)\dot{\alpha}^2, \label{E_dissipation}
\end{align}
where $\eta$ is the effective dynamic viscosity with a typical order of 1~Pa$\cdot$s.
In order to obtain the equation governing the wrapping dynamics of the active particle, first of all we construct a Rayleighian $\mathcal{R}=\dot{E}_{\mathrm{tot}}+\Phi$, with $\dot{E}_{\mathrm{tot}}$ the time derivative of the free energy of the system given by
\begin{align}
\dot{E}_{\mathrm{tot}}=&\biggl\{\frac{\kappa e^2}{a^2}\frac{[2+(e^2-1)\sin^2\alpha]^2}{[1+(e^2-1)\sin^2\alpha]^3}+  \notag\\
&2\sigma\biggl(1-\frac{\cos\alpha}{\sqrt{\cos^2\alpha+e^2\sin^2\alpha}}\biggr)-2\omega\biggr\}\times  \notag\\
&\pi a^2 \sin\alpha\sqrt{\cos^2\alpha+e^2\sin^2\alpha}\dot{\alpha}-fae\sin\alpha\dot{\alpha}, \label{dEdalpha}
\end{align}
and $\Phi$ the energy dissipation function. Minimizing $\mathcal{R}$ with respect to $\dot{\alpha}$ following the Onsager variational principle, i.e., $\partial\mathcal{R}/\partial\dot{\alpha}=0$, we obtain the cellular uptake dynamics equation
\begin{align}
\dot{\alpha}=&\frac{1}{\eta a\sqrt{\cos^2\alpha+e^2\sin^2\alpha}}\biggl\{\omega+ \frac{fe}{2\pi a\sqrt{\cos^2\alpha+e^2\sin^2\alpha}} \notag\\
&-\frac{\kappa e^2}{2a^2}\frac{[2+(e^2-1)\sin^2\alpha]^2}{[1+(e^2-1)\sin^2\alpha]^3} \notag\\
&-\sigma\biggl(1-\frac{\cos\alpha}{\sqrt{\cos^2\alpha+e^2\sin^2\alpha}}\biggr)\biggl\}, \label{dynamicequation}
\end{align}
for a nonspherical active particle. For spherical particles $a=b=R$, the above equation reduces to Eq.~(2) in Ref.~\cite{F.Frey2019} if $f=0$.

 A detailed theoretical analysis of Eq.~(\ref{dynamicequation}) shows that there exist two types of critical conditions for a full uptake to occur. One is governed by $\dot{\alpha}\bigl|_{\alpha=\pi}=0$, or
\begin{align}
\omega+\frac{f}{2\pi a}e-\frac{2\kappa}{a^2}e^2-2\sigma=0, \label{e_c}
\end{align}
corresponding to a second-order wrapping transition for prolate particles and spherical particles. For a spherical particle, Eq.~(\ref{e_c}) reduces to a critical radius $R_c=[-f+\sqrt{f^2+32\pi^2\kappa(\omega-2\sigma)}]/[4\pi(\omega-2\sigma)]$ when $\omega\neq2\sigma$. The other is given by $\dot{\alpha}\bigl|_{\alpha=\pi/2+B/(2C)}=0$ with $B=\sigma/e$, and
\begin{align}
C=\frac{1-e^2}{2e^2}\biggl(\sigma-\omega-\frac{f}{\pi a}\biggr)+\frac{\kappa(3e^2+7)}{4a^2e^6},\nonumber
\end{align}
leading to a critical condition
\begin{align}
\omega+\frac{f}{2\pi a}-\frac{\kappa (1+e^2)^2}{2a^2e^4}-\sigma-\frac{B^2}{4C}=0, \label{large tension}
\end{align}
for a first-order wrapping transition from a partial wrapping to a full uptake. If the surface tension satisfies $(2e^2-1)\sigma\ll \kappa(3e^2+7)/(a^2e^4)-2(1-e^2)[\omega+f/(\pi a)]$, the above condition reduces to
\begin{align}
\omega+\frac{f}{2\pi a}-\frac{\kappa (1+e^2)^2}{2a^2e^4}-\sigma=0. \label{e_cl}
\end{align}

To investigate the wrapping dynamics of the active particle, we first solve Eq.~(\ref{dynamicequation}) numerically and analyze the evolution of the wrapping angle $\alpha/\pi$. For an oblate spheroid, the engulfing angle (black curves) as shown in Fig.~\ref{tVsR}(a) exhibits a plateau around $\alpha=\pi/2$, indicating that there exists a stable partial wrapping state at this point. This can be reflected by its corresponding low wrapping velocity (blue curve) as the oblate particle has to cross over an energy barrier before it is fully wrapped, corresponding to a first-order wrapping transition. Figures~\ref{tVsR}(b) and~\ref{tVsR}(c) show the similar wrapping behavior with a monotonic increase of wrapping angle to a completely wrapped state ($\alpha/\pi=1$) for both the prolate spheroid and the spherical particle. The wrapping velocity for the spherical particle shows a significant and monotonic drop off after the particle is initially internalized at a fast speed, until it slows down and is fully wrapped by the fluidic film [see Fig.~\ref{tVsR}(b)].
While for the prolate spheroid, the engulfing speed increases against time in the beginning, followed by a dramatic decrease, and finally levels off to zero, terminating at a full wrapping state ($\alpha/\pi=1$) [see Fig.~\ref{tVsR}(c)].

\begin{figure}[htp]
  \includegraphics[width=\linewidth,keepaspectratio]{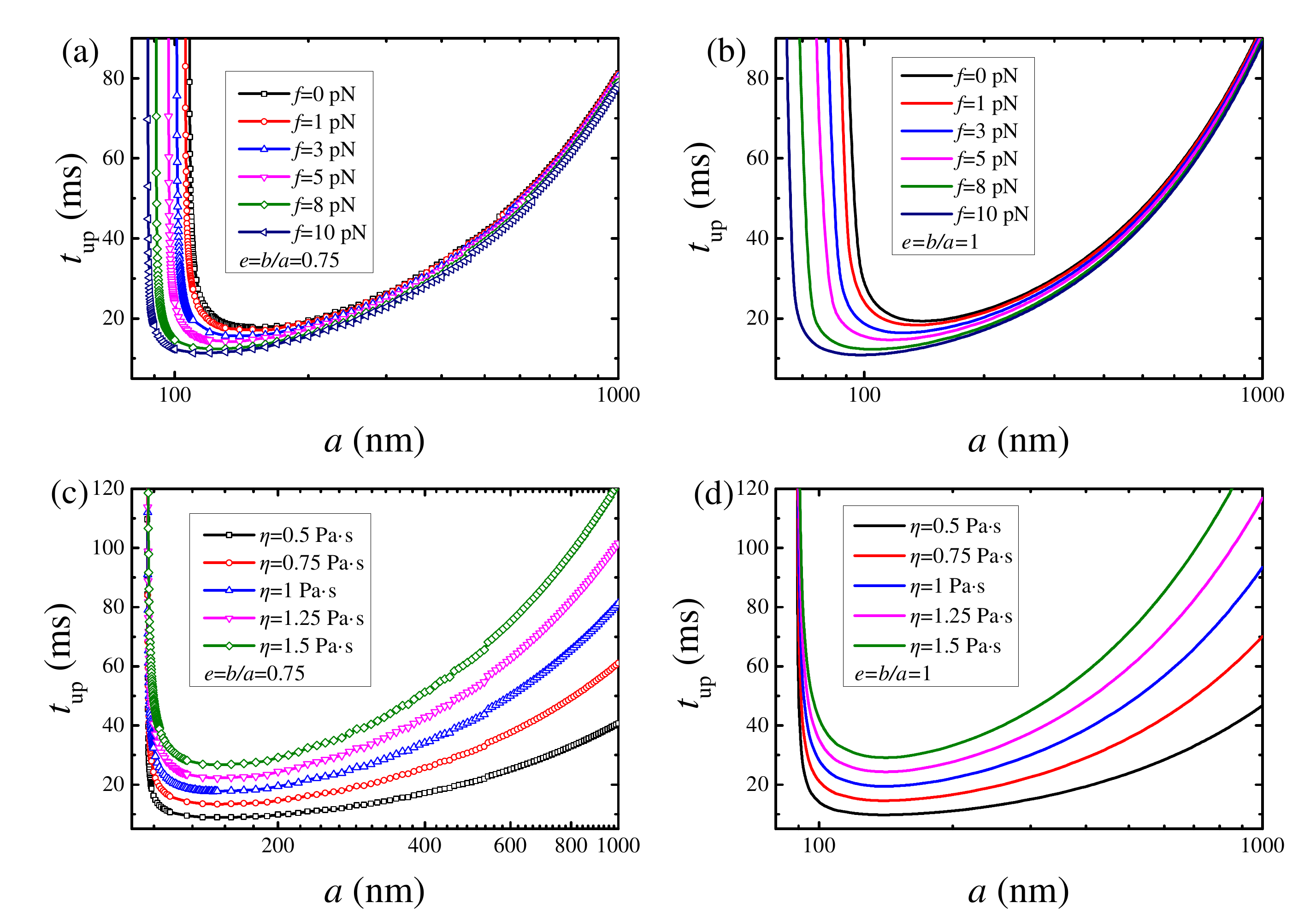}
  \caption{(Color online) Wrapping times as functions of particle size for (a) oblate particle, and (b) spherical particle with different active forces.  Uptake time as a function of particle size for (c) oblate particle, and (d) spherical particle with different effective dynamic viscosities. Parameter values (if not varied) are fixed at $\kappa=25~k_BT$, $\omega=0.044~{\rm mJ/m^2}$, $\sigma=0.9\times10^{-5}~{\rm N/m}$, and $f=0~{\rm pN}$, $\eta=1~{\rm Pa\cdot s}$. \label{t Vs b}}
\end{figure}

To systematically study how the active force and the aspect ratio of the particle, the effective dynamic viscosity, and the membrane properties affect the uptake dynamics of the active particle at cell membrane, we numerically calculate the wrapping time by changing these parameter values. Figure~\ref{t Vs b} shows the wrapping time as a function of particle size for different active forces and dynamic viscosities.
For fixed $f$ and $\eta$, the curves of uptake time share a similar feature that with the increase of the particle size, $t_{\rm up}$ drops very quickly before it bounces back mildly.
Meanwhile, the full wrapping regimes in the parameter space are also apparently broadened with the increase of the active force and the effective dynamic viscosity.
Figures~\ref{t Vs b}(a) and~\ref{t Vs b}(b) compare the resulting wrapping times of the oblate and the spherical particle for different values of active force. For the oblate particle, we adjust the length of the symmetry axis $a$ by keeping the aspect ratio unchanged. Both ellipsoid and spherical particles share a similar behavior that there exists a critical particle size beyond which a full uptake occurs. Interestingly, the calculation results demonstrate that particles with larger active forces are uptaken faster than those with smaller forces, indicating that particle activity facilitates the uptake process. This conclusion is in line with the simulation predictions in Ref.~\cite{P.Chen2020}, where the authors claimed that the uptake efficiency can be enhanced with the increase of P\'{e}clet number quantifying the strength of active force.

In order to probe how effectively the dynamic viscosity affect the wrapping dynamics, we plot the wrapping time for an oblate particle and a spherical particle against particle size with different effective dynamic viscosities, as shown in Figs.~\ref{t Vs b}(c) and~\ref{t Vs b}(d), respectively.
It is found that decreasing the effective dynamic viscosity $\eta$ clearly decreases the threshold particle size for the occurrence of the complete uptake. This is due to the fact that large viscosity increases the friction, which correspondingly inhibits the uptake process and hence leads to longer uptake time. Therefore, to achieve faster (slower) wrapping process, enhancing (weakening) the particle activity and reducing (raising) the effective dynamic viscosity might be an effective option.

\begin{figure}[htp]
  \includegraphics[width=\linewidth,keepaspectratio]{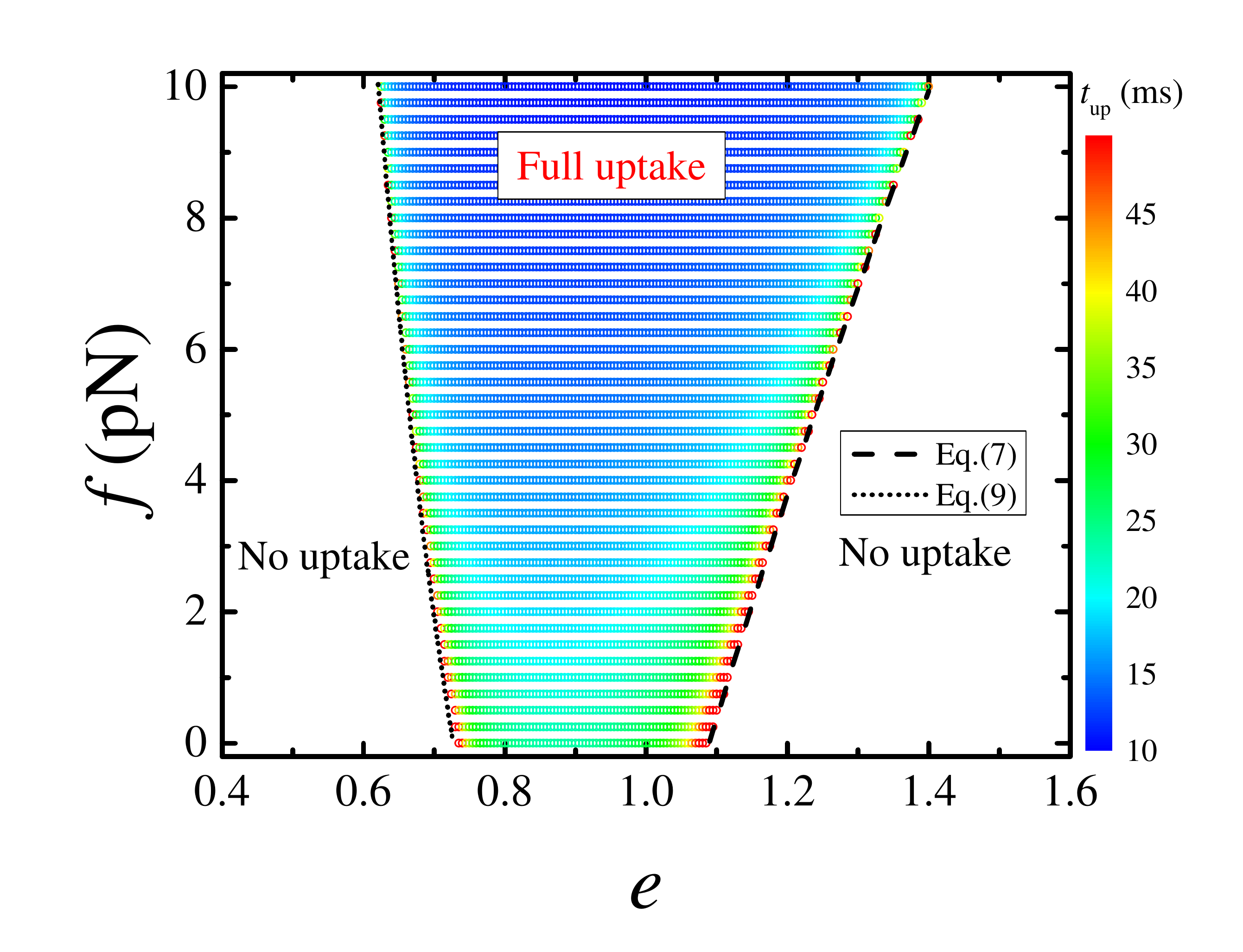}
  \caption{(Color online) A two-dimensional phase diagram on the ($f-e$) plane characterizes the interrelated effects of active force and particle aspect ratio on the uptake time of wrapping process, where the bending rigidity, the adhesion energy density, the membrane tension, and the effective dynamic viscosity are set as $25~k_BT$, $0.044~{\rm mJ/m^2}$, $0.9\times10^{-5}~{\rm N/m}$, and $1~{\rm Pa\cdot s}$, respectively. The volume of the particle is fixed at $V=4\pi ea^3/3=4\pi R_0^3/3$ with $R_0=100~{\rm nm}$. \label{f Vs e}}
\end{figure}

Pathogens and viruses come in many different shapes~\cite{C.Hulo2011}, but the most frequent occurrence is ellipsoid.
Therefore, here we focus our discussion on ellipsoid with its shape characterized by aspect ratio.
Furthermore, to gain more insight into the interrelated influence of the active force $f$ and the particle aspect ratio $e$ on the wrapping time, we construct a phase diagram for the wrapping time in the $f-e$ space under the condition of fixed particle volume, as shown in Fig.~\ref{f Vs e}.
On one hand, a comparison of the wrapping time between the prolate ellipsoidal particle ($e>1$) and the spherical particle ($e=1$) shows that an prolate ellipsoidal particle is taken up slower than a spherical one. However, upon decreasing the aspect ratio, the uptake time for the active oblate ellipsoidal particles ($e<1$) displays a nonmonotonic feature by decreasing to a minimum value first and then bouncing back gradually, in stark contrast to the monotonic dependence for active prolate particles with an aspect ratio of $e>1$. On the other hand, Figure~\ref{f Vs e} also verifies that enhancing the activity gives rise to the wrapping efficiency (with a decrease of $t_{\rm up}$). Therefore, the wrapping efficiency can be regulated by tuning the aspect ratio and the activity of particles.
To determine the boundaries separating the full uptake and no uptake regimes, we plotted two boundary curves based on Eq.~(\ref{e_c}) [see the dash curve in Fig.~\ref{f Vs e}] and Eq.~(\ref{e_cl}) [see the dot curve in Fig.~\ref{f Vs e}], respectively, which match the numerical calculations very well. The consistence once again indicates that there exist two ways for a nonspherical particle to reach full wrapping, i.e. a continuous uptake for prolate particles and a snapthrough uptake for oblate particles. Such a conclusion is in agreement with the simulations done by Khosravanizadeh \textit{et. al}~\cite{Khosravanizadeh2022}, who demonstrated that the oblate ellipsoidal particles exhibit discontinuous wrapping phase transition from partial wrapping to full wrapping during the uptake process, while the prolate ellipsoidal particles show a continuous wrapping transition behavior.

\begin{figure}[htp]
  \includegraphics[width=\linewidth,keepaspectratio]{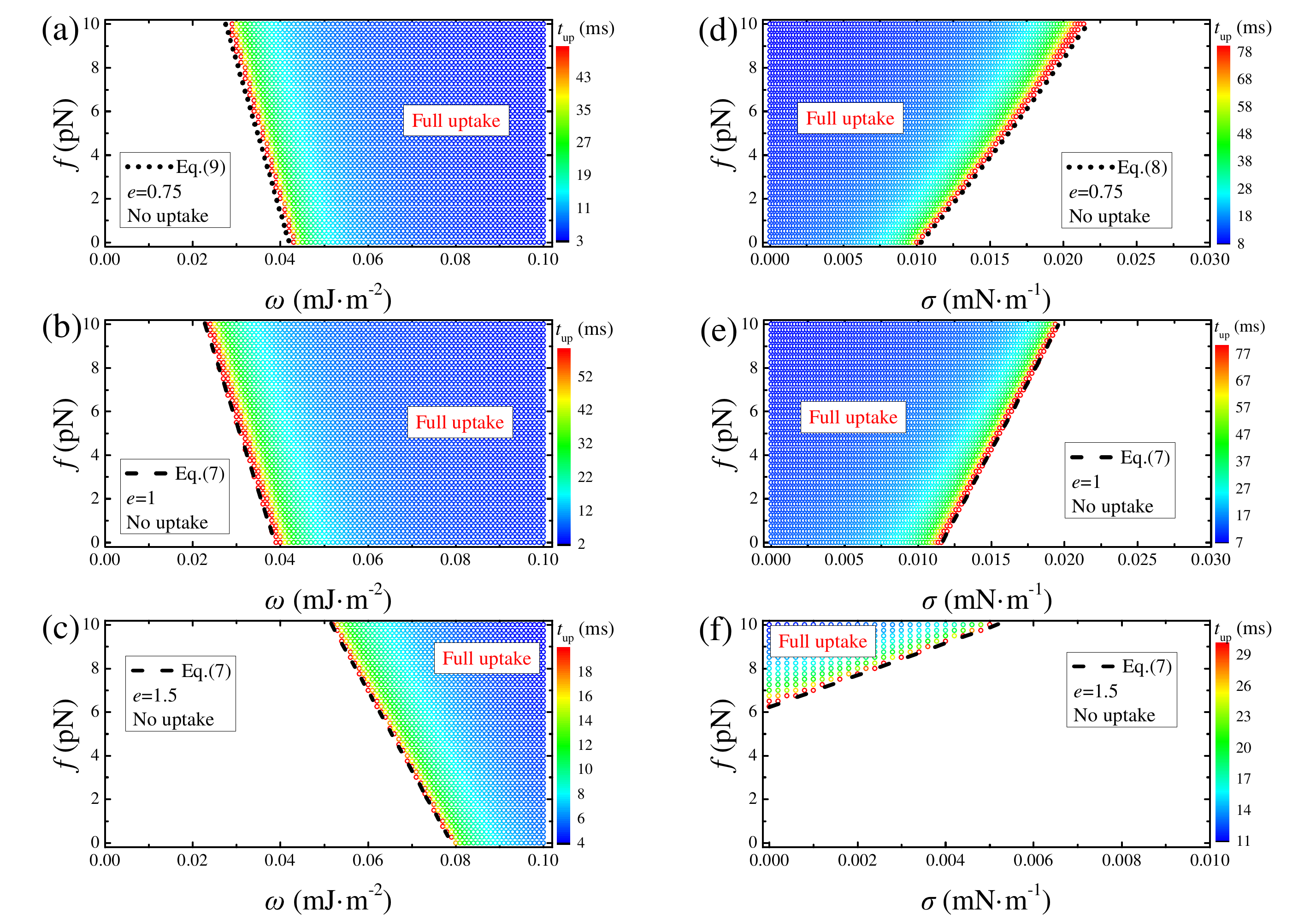}
  \caption{(Color online) Wrapping time $t_{\rm up}$ as a function of $f$ and $\omega$ for (a) oblate ellipsoidal, (b) spherical, and (c) prolate ellipsoidal particles at equal particle volume. Uptake times as a function of active force $f$ and membrane tension $\sigma$ for (d) oblate ellipsoidal, (e) spherical, and (f) prolate ellipsoidal particles at equal particle volume. Parameter values (if not varied) are fixed at $\kappa=25~k_BT$, $\omega=0.044~{\rm mJ/m^2}$, $\sigma=0.9\times10^{-5}~{\rm N/m}$, and $\eta=1~{\rm Pa\cdot s}$. The volume of the particle is fixed at $V=4\pi ea^3/3=4\pi R_0^3/3$ with $R_0=100~{\rm nm}$.  \label{t Vs w}}
\end{figure}

In order to gain more insights into the effects of the membrane properties on the wrapping time, we explore the wrapping dynamics of an active oblate particle, a prolate ellipsoidal one, and a spherical one with different adhesion energy densities and membrane tensions, as shown in Fig.~\ref{t Vs w}.
The colored contour maps of $t_{\rm up}$ on the $f-\omega$ [see Figs.~\ref{t Vs w}(a), \ref{t Vs w}(b), and \ref{t Vs w}(c)] and $f-\sigma$ [see Figs.~\ref{t Vs w}(d), \ref{t Vs w}(e), and \ref{t Vs w}(f)] planes show that given an aspect ratio and a fixed active force, a higher (lower) wrapping efficiency can be achieved under a stronger (weaker) adhesion force or a lower (higher) membrane tension, i.e. higher adhesion and looser membrane leads to faster wrapping. Such a conclusion can be supported by experimental observations~\cite{H.T.Spanke2022} reported recently that higher adhesion leads to faster wrapping, and simulation results in Ref.~\cite{F.Frey2019} that the uptake time is strongly decreased for higher adhesion and looser membrane. Here it is important to note that the boundaries of the full uptake (read lines) can be very well captured by the critical conditions Eqs~(\ref{e_c}), (\ref{large tension}) and (\ref{e_cl}) theoretically derived using our model.

\begin{figure}[htp]
  \includegraphics[width=\linewidth,keepaspectratio]{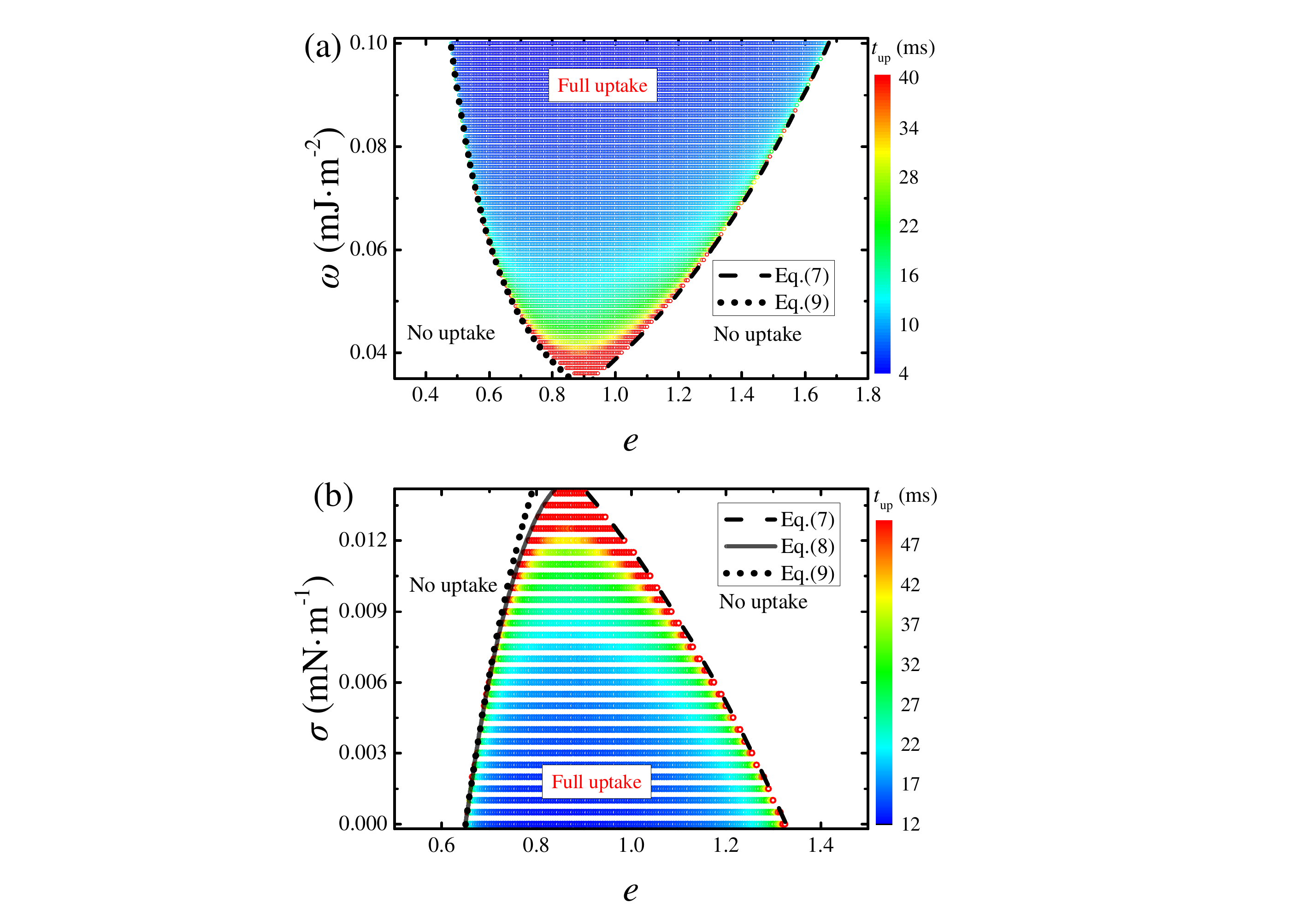}
  \caption{(Color online) Two-dimensional wrapping time phase diagrams in the projection planes of (a) $\omega$ and $e$ and (b) $\sigma$ and $e$, where the bending rigidity, the adhesion energy density, the membrane tension, the active force and the effective dynamic viscosity (if not varied) are set as $25~k_BT$, $0.044~{\rm mJ/m^2}$, $0.9\times10^{-5}~{\rm N/m}$, $f=0~{\rm pN}$ and $1~{\rm Pa\cdot s}$, respectively. The volume of the particle is fixed at $V=4\pi ea^3/3=4\pi R_0^3/3$ with $R_0=100~{\rm nm}$. \label{w Vs e}}
\end{figure}

Finally we discuss the influence of the membrane properties and the aspect ratio of particle to the wrapping process. Figures~\ref{w Vs e}(a) and~\ref{w Vs e}(b) show the two-dimensional phase diagrams of wrapping time in the parametric planes of $\omega-e$ and $\sigma-e$.
What is striking is that, for the oblate ellipsoidal particles, the uptake time nonmonotonically depends on $e$ within the scale of $\omega$ or $\sigma$, in comparison to the monotonic dependence of the prolate ellipsoidal particles. Tuning the particle's aspect ratio to an optimal value enables one to obtain a minimal wrapping time.
In addition, Figure~\ref{w Vs e} also confirms that the uptake efficiency can be improved (abated) by increasing (decreasing) the adhesion energy density, and reducing (raising) the membrane tension.
The dash and solid curves based analytical results Eq.~(\ref{e_c}) and Eq.~(\ref{large tension}) coincide with the boundaries obtained from numerical calculations.
It has been shown, by providing a stochastic model to study the kinetics of particle wrapping by a vesicle, that increasing the attraction strength between the particle and vesicle causes the improvement of uptake rate~\cite{S.Mirigian2013}. This is again a result in agreement with our present conclusions.

Consequently, we argue that the wrapping time decreases with the increase of the active force; the higher the viscosity, the longer the wrapping time; strong adhesion and low tension improve the wrapping efficiency. The reason stems from that the wrapping process is largely controlled by the competition among the three types of energy: the elastic energy (consisting of bending energy and tension energy), the adhesive energy, and the work done by the active force.
Reducing wrapping time can be realized based on the condition that the adhesion energy and the work done by the active force is sufficient to overcome the energy barrier, namely, the sum of the elastic energy and the viscous dissipation. In the presence of active force, the work done by the active force reduces the free energy and as a result change the uptake force. The positive correlation between active force and uptake force gives rise to the positive dependence of wrapping time on the active force.
According to Eq.~(\ref{dynamicequation}), the adhesion energy driving the wrapping process is positively proportional to the adhesion energy density and negatively proportional to the membrane tension.
As a result, increasing the adhesion energy density corresponds to increasing the driving force for wrapping, decreasing the membrane tension decreases the energy penalty for uptake, indicating a decreases of the wrapping time. As for the effective dynamic viscosity, an increase of it means that the induced energy dissipation requires more adhesion energy and work done by the active force to compensate,  which leads to an increase of wrapping time.

In summary, we propose a theoretical model to investigate the wrapping dynamics of a nonspherical active particle by a lipid plasma membrane, by taking into account the influence of the active force, the particle shape, the effective dynamic viscosity, and the membrane properties (including the adhesion energy density and the membrane tension).
The wrapping equation for the active particle, which quantitatively couples the elastic deformation of the membrane, the work done by the active particle, and the energy dissipation, is derived by using Onsager variational principle. Two critical conditions, one for the continuous full uptake of prolate particles and the other for snapthrough full wrapping of oblate particles, are obtained theoretically. Our results reveal that enhancing activity (active force), reducing effective dynamic viscosity, increasing adhesion energy density, and decreasing membrane tension, can significantly improve the wrapping efficiency for the self-propelled particles.
Intriguingly, with an increase of aspect ratio, the wrapping time for oblate ellipsoidal particles exhibits a nonmonotonic dependence, in stark contrast to the monotonic dependence for prolate ellipsoidal particles.
Therefore, the wrapping time can be manipulated by changing the activity and the aspect ratio of the particles, the effective dynamic viscosity, and the properties of the membrane, such as adhesion energy density and membrane tension.
All these findings may not only shed light on the influence of the activity and the aspect ratio of particles, the viscosity, and the properties of membrane on the dynamic behaviors of wrapping process, but also provide guidelines to improve the efficiency of active particle-based drug delivery systems.

\section{ACKNOWLEDGMENTS}
We acknowledge financial support from National Natural Science Foundation of China under Grant Nos.12147142, 11974292, 12174323, and 1200040838, and 111 project B16029.

\end{document}